\documentclass[aps,showpacs,showkeys,preprint]{revtex4}
\usepackage{epsbox}
\usepackage[dvips]{color}

\begin{document}
\preprint{KOBE-TH-03-04}

\title{Escape from washing out of baryon number in a two-zero-texture general Zee model
 compatible with the large mixing angle MSW solution }



\author{~K.~Hasegawa}
\email[]{kouhei@phys.sci.kobe-u.ac.jp}
\author{~C.~S.~Lim}
\email[]{lim@phys.sci.kobe-u.ac.jp}
\author{~K.~Ogure}
\email[]{ogure@phys.sci.kobe-u.ac.jp}
\affiliation{Department of Physics, Kobe University, Rokkodaicho 1-1,
Nada ward, Kobe 657-8501, Japan}


\date{\today}

\begin{abstract}
We propose a two-zero-texture general Zee model, compatible with 
the large mixing angle Mikheyev-Smirnov-Wolfenstein solution.
The washing out of the baryon number does not occur in
this model for an adequate parameter range.  We check the consistency
of a model with the constraints coming from flavor changing neutral
current processes, the recent cosmic microwave background observation,
and the Z-burst scenario.
\end{abstract}



\maketitle

\section{Introduction\label{intro}}

Recent neutrino oscillation experiments reveal that neutrinos have
small but finite mass differences \cite{S-K,SNO,kam}. The precise
observation of the cosmic microwave background (CMB) shows that not
only the mass differences, but also the masses themselves, are
small \cite{Wmap1,Wmap2,Wmap3,Wmap4,Han,Elg}.  The smallness of the neutrino
mass is naturally explained in the seesaw mechanism \cite{Yan,Yan2,Yan3}
with right-handed neutrinos and in the Zee model with radiative mass
generation \cite{Zee}.  In the present paper, we focus on the Zee model, 
which has many predictions for low energy physics.  This model has two
Higgs doublets and a singlet Higgs scalar field (the Zee singlet), in
addition to the fields in the standard model. In the 
original Zee model, both of the two Higgs doublets couple to the
ordinary matter fields.  Wolfenstein, however, imposed a discrete symmetry
 on the Zee model so that only one of the two Higgs doublets
couples to the ordinary matter fields.  As a result, dangerous
processes through the flavor changing neutral current (FCNCs ) are easily
avoided in this model, and this simplified model has been mainly considered
so far \cite{Wol}. We call this model the restricted Zee
model (RZM) \cite{Gri}. One of the attractive points of the Zee model in
this version is that the bilarge mixing angle pattern, with respect
to the solar and atmospheric neutrino oscillations, is naturally
explained.  This attractive feature, ironically, has forced the Zee
model into a difficult position very recently.  The predicted large
angles must be just around $\pi/4$ \cite{Jarl,Fra1,Koide} and are
incompatible with the large mixing angle (LMA) 
Mikheyev-Smirnov-Wolfenstein (MSW) solution with
$\theta_{\odot}\sim\pi/5.6$, while the KamLAND experiment reports that
only the LMA MSW solution is allowed \cite{kam}.  Several ideas have been
proposed to escape from this difficulty: introducing a fourth
neutrino \cite{Roy,Pel}, introducing a doubly charged Higgs scalar
field \cite{Fra2}, introducing triplet Higgs scalars \cite{Sch,Yas,Fra4},
and allowing both of the two Higgs doublets to couple to the ordinary
matter fields.

Aside from the difficulty concerning the solar mixing angle, we have
another issue with this model in cosmology.  Since the lepton number is
violated in this model, the primordial baryon number is washed out
through the sphaleron process at quite a low temperature \cite{Ma,Sar}.
This model therefore seems to be unable to exist together with
baryogenesis scenarios in which the baryon number is produced at a high
temperature.  However, it has been pointed out that there exists an
approximately conserved number $L^{'}\equiv L_e-L_{\mu}-L_{\tau}$ for
the low mass, low probability (LOW) and vacuum oscillation (VO)
 solutions of the solar neutrino problem in the RZM, and complete baryon number washing out does not
occur \cite{Haba}.

In the present paper, we focus on the original Zee
 model in which both of the two Higgs doublets couple to the ordinary
matter fields. We call this model the general Zee
model (GZM) in order to clearly distinguish it from the
RZM \cite{Gri}.  Although this model was proposed earlier, it
has not been extensively discussed, as we mentioned above.  We shed
light on the GZM once again and find not only that the two
phenomenological problems above can be avoided
\footnote{We note that the constraint from the baryon number washing
out is also avoided by baryogenesis at a temperature lower than the
scale of the singlet Higgs boson mass, e.g., electroweak
baryogenesis \cite{Kuz,Coh} or low temperature
leptogenesis \cite{Ham}.  The order of the electroweak phase transition
can be altered due to the additional Higgs boson scalar
field  \cite{Valle-electro}.  On the other hand, low temperature
leptogenesis is impossible in the model we propose in this paper
because of the absence of CP violation and an adequate
nonequilibrium.  Some additional extensions are needed for it.}
 but also that the GZM
can be consistent with experiments involving FCNCs. This model is
attractive in the sense that the extension from the RZM,
 of which properties have been extensively discussed,
is minimal.  Although the extension is minimal,  many undetermined parameters arise  
and the investigation in the full parameter region is very difficult.  We then constrain the structure of the
neutrino mass matrix obtained from the Zee model.  Two constraints for
the neutrino mass matrix have been proposed: imposing a texture to
have two independent zeros in the base, where the masses of the charged
leptons are diagonalized \cite{Fra3}, and requiring the mass matrix to
have a vanishing determinant \cite{Bra}.  While the vanishing
determinant constraint is good in that the condition does not depend
on the weak base we take, and it is suitable for an investigation into
the Zee model with a doubly charged  Higgs singlet \cite{Zee2,Bab}, it
does not fit the structure of the mass matrix obtained from the Zee
model considered here.  We therefore employ the two-independent-zero-texture
 constraint, of which phenomenological aspects were investigated
in Refs. \cite{Xin1,Xin2,Des}.  The suitable two-zero-texture mass
matrix for the Zee model considered here is the case C defined in
Ref. \cite{Fra3},
\begin{eqnarray}
{\cal M}_{\nu}=\left(\begin{array}{rrr}
    \times   &   \times   & \times    \\
    \times   &   0   &  \times     \\
   \times   &   \times   &  0     
	\end{array}\right).
\end{eqnarray}
The main aim of the present paper is to search for the simplest GZM
with this structure of the mass matrix, which is compatible with the
LMA MSW solution of the solar neutrino problem and free from the
baryon number washing out \footnote{We do not consider how the
primordial baryon number is produced.  Baryogenesis in an extended
version of the Zee model is discussed in Ref. \cite{Ham}.}.  It turns
out that such a GZM can be constructed by minimal extension from the
RZM, i.e., by adding only one coupling constant to the RZM.  We also
examine whether this model is allowed from the viewpoint of
experiments: $\mu$($\tau$) decay rate, $\mu(\tau) \rightarrow e
\gamma$ \cite{Gri} and neutrinoless double beta decay \cite{Che}.
We briefly consider the constraints from the recent CMB
observation \cite{Wmap1,Wmap2,Wmap3,Wmap4} and the possibility of the
Z-burst scenario in this model \cite{Wei,Far}.

The outline of this paper is as follows: In Sec. II, we first
analyze the structure of the neutrino mass matrix that is compatible
with SNO/S-K and CHOOZ data. We next derive the Majorana neutrino mass
matrix that is induced through the radiative correction in the GZM.
The parameters of the GZM are determined by comparing the structure of
the neutrino mass matrix obtained from the experiments with that
obtained from the GZM.  In Sec. III, we investigate the cosmological
implication of the GZM.  From the viewpoint of baryogenesis, there
must be some combinations of the lepton numbers that avoid the washing
out of the primordial baryon number.  We first examine what
combinations of lepton numbers are conserved in each type of GZM.
 We next obtain a range of parameters in which the baryon number
is not washed out. We finally calculate the resulting lepton and
baryon number.  Section IV is devoted to a summary.

\section{The GZM mass matrix compatible with the LMA-MSW solution \label{II}}
In the present section, we review the structure of the neutrino mass
matrix arising in the GZM and determine the parameters of the GZM using
the neutrino oscillation data.  We further examine whether the GZM is
consistent with the constraints from other phenomenological
experiments.
\subsection{Neutrino mass matrix that satisfies SNO/S-K and CHOOZ experimental data \label{A}}

 The Majorana neutrino mass term is written as  
\begin{eqnarray}
{\cal L}_{Majorana}=\frac{1}{2} \vec{\bar{\nu}}_{L}^{c} {\cal M}_{\nu} \vec{\nu}_{L}  + \mbox{H.c.}
 \ \ \mbox{for} \ \ 
 \vec{\nu}_{L}\equiv \left(\begin{array}{c}
    	\nu_{e}\\   
       \nu_{\mu}  \\ 
         \nu_{\tau}
	\end{array}\right)_{L},
\end{eqnarray}
where ${\cal M}_{\nu}$ is a complex valued symmetric matrix. Here, the
weak eigenstate $\vec{\nu}_{L}$ is taken to diagonalize the mass matrix
of the charged leptons.  Using the Maki-Nakagawa-Sakata (MNS) matrix U,
we can diagonalize ${\cal M}_{\nu}$ to $\hat{\cal M}_{\nu}$,
\begin{eqnarray}
 \hat{\cal M}_{\nu} &=& U^{T}{\cal M}_{\nu}U,  \\
 \vec{\nu}_{L}^{m} &=& U^{\dagger} \vec{\nu}_{L}\equiv \left(\begin{array}{c}
    	\nu_{1}\\   
       \nu_{2}  \\ 
         \nu_{3}
	\end{array}\right)_{L},
\end{eqnarray}
where $\vec{\nu}_{L}^{m}$ is in the neutrino mass eigenstate.  We write
 a diagonal matrix with a caret hereafter.  The
 Lagrangian density is written in this base,
\begin{eqnarray}
{\cal L}_{Majorana}=\frac{1}{2} (\vec{\bar{\nu}}_{L}^{m})^{c} \hat{\cal M}_{\nu} \vec{\nu}_{L}^{m} + 
\mbox{H.c.}
\end{eqnarray}
We assume that ${\cal M}_{\nu}$ is real and the MNS matrix is orthogonal
for simplicity.  Under these assumptions, we can parametrize the
orthogonal matrix U as
\begin{eqnarray}
 U &\equiv&  
   \left(\begin{array}{ccc}
    1   &   0   & 0    \\   
    0   &   c_{23}   &  s_{23}     \\ 
   0   &   -s_{23}    &  c_{23}     
	\end{array}\right)
 \left(\begin{array}{ccc}
    c_{13}   &   0   & s_{13}    \\   
    0   &   1   &  0      \\ 
   -s_{13}   &   0    &  c_{13}     
	\end{array}\right)
 \left(\begin{array}{ccc}
    c_{12}   &   s_{12}   & 0    \\   
    -s_{12}   &   c_{12}  &  0      \\ 
   0  &   0    &  1     
	\end{array}\right)  \nonumber \\
	&=& R_{23}(-\theta_{23})R_{13}(\theta_{13})R_{12}(-\theta_{12}), 
\end{eqnarray}
where $c_{ij} \equiv \cos \theta_{ij}$ and $s_{ij} \equiv \sin \theta_{ij}$. 

\par We summarize the current results of neutrino experiments to make
clear what type of neutrino mass matrix is allowed.  The S-K
experiment shows that there exist a mass squared difference
$\Delta_{a}$ and a mixing angle $\theta_{atm}$ in order to explain the
atmospheric neutrino oscillation \cite{S-K,S-K2,Valle-Gon} of
\begin{eqnarray}
 1.6 \times 10^{-3} \ \mbox{eV}^2 < \Delta_{a} < 4.0 \times 10^{-3} \ \mbox{eV}^2, \ \ 
0.88 < \sin^2{2\theta_{atm}} \le 1.0  \ (90 \% \ \mbox{C.L.}),   \label{sno}
\end{eqnarray}
with the best fit values $\Delta_{a}=2.5 \times 10^{-3} \ \mbox{eV}^2$
and $\sin^2{2\theta_{atm}}=1.00$. The global analysis
of the first results of KamLAND combined with existing data from
solar neutrino experiments shows that there exist a mass squared
difference $\Delta_{s}$ and a mixing angle $\theta_{\odot}$ for
explaining the solar neutrino oscillation
\cite{SNO,kam,SNO2,kam2,Valle-kam} of
\begin{eqnarray}
5.1 \times 10^{-5}\  \mbox{eV}^2 < \Delta_{s} < 9.7 \times 10^{-5} \ \mbox{eV}^2&,& \  \ 
1.2 \times 10^{-4} \ \mbox{eV}^2 < \Delta_{s} < 1.9 \times 10^{-4} \ \mbox{eV}^2, \label{two}
\\ 0.29 < \tan^2{\theta_{\odot}} <0.86 && (3 \sigma \ \mbox{level}), \label{non}
\end{eqnarray}
with the best fit values, $\Delta_{s}=6.9 \times 10^{-5}
\ \mbox{eV}^2$ and $\tan^2{\theta_{\odot}}=0.46$ \cite{Valle-kam}. This analysis
concludes with three remarks: first, the original LMA region is narrowed
and is separated into two islands (\ref{two});  second, the LMA
 is the unique solution to 4$\sigma$ level; thirdly, the maximal
solar neutrino mixing angle is excluded at the 3$\sigma$ level (\ref{non}).
 These data are rewritten using the mass squared
difference ratio R and the deviation of the solar neutrino angle from
$\pi/4$, $\theta_{\odot} \equiv \pi/4-\epsilon_{3}$, as
\begin{eqnarray}
1.3 \times 10^{-2} < R \equiv \frac{\Delta_{s}}{\Delta_{a}} < 1.2 \times 10^{-1}, \ \  
0.038 <\epsilon_{3}<0.29 \ (3 \sigma \ \mbox{level}), \label{R3}
\end{eqnarray}
with the best fit values $R$ = $2.8 \times 10^{-2}$ and
$\epsilon_{3}=0.19$. The CHOOZ experiment
shows the upper bounds for the mixing angle $\theta_{13}$ \cite{Chooz} to be
\begin{eqnarray}
\sin^2{2\theta_{13}} \le 0.1. \label{CH}
\end{eqnarray}

\par
Since we assume that the mass matrix ${\cal M}_{\nu}$ is real, we can generally parametrize it as 
\begin{eqnarray}
{\cal M}_{\nu}&=&\left(\begin{array}{rrr}
    a   &   +b   & -b    \\
    +b   &   c   &  d     \\
   -b   &   d   &  c     
	\end{array}\right)
	+\eta \left(\begin{array}{rrr}
    0   &   1   & 1    \\
    1   &   0   &  0     \\
   1   &   0   &  0     
	\end{array}\right)
	+\xi \left(\begin{array}{ccc}
    0   &    0   & 0    \\
    0   &   1   &  0     \\
    0 &   0   &  -1     
	\end{array}\right)\\
	&\equiv& {\cal M}_{\nu}^{0} +{\cal M}_{\nu}^{\eta} + {\cal M}_{\nu}^{\xi}. \label{nm}
\end{eqnarray}

Using the orthogonal matrix
\begin{eqnarray}
 U_{0} \equiv R_{23}(-\frac{\pi}{4})R_{13}(0)R_{12}(-\frac{\pi}{4}+\epsilon_{3})
 \ \ \mbox{with}\ \ \mbox{tan2}\epsilon_{3}=\frac{c-a-d}{2\sqrt{2}b},
\end{eqnarray}
the first matrix ${\cal M}_{\nu}^{0}$ in Eq. (\ref{nm}) is diagonalized as follows \cite{Gri}:
\begin{eqnarray}
   U^{T}_{0}{\cal M}_{\nu}^{0}U_{0}&=&R_{12}(-\epsilon_{3})
\left(\begin{array}{ccc}
    -\sqrt{2}b+\frac{1}{2}(a+c-d)   &  \frac{1}{2}(a-c+d)   & 0    \\
    \frac{1}{2}(a-c+d)              &   \sqrt{2}b+\frac{1}{2}(a+c-d)   &  0     \\ 
   0   &   0   &  c+d    
	\end{array}\right)R_{12}(\epsilon_{3}) \nonumber \\
	&=& \left(\begin{array}{ccc}
    \frac{a+c-d+\sqrt{(a-c+d)^2+8b^2}}{2}  & 0    & 0    \\
      0        &  \frac{a+c-d-\sqrt{(a-c+d)^2+8b^2}}{2}  &  0     \\ 
   0   &   0   &  c+d
	\end{array}\right).
\end{eqnarray}
According to the data (\ref{sno}) and (\ref{CH}), the deviations of the
angles $\theta_{23}$ and $\theta_{13}$ from the values
$\theta_{23}=\pi/4$ and $\theta_{13}=0$ are small, and the
influence of $\eta$ and $\xi$ is expected to be treated as a small
perturbation.  Defining these deviations as
$\epsilon_{1}\equiv \pi/4-\theta_{23}$ and $\epsilon_{2}\equiv
\theta_{13}$, we find that they should lie in the range 
$0<\epsilon_{1}<0.18$ and $0 <\epsilon_{2}<0.16$.  We obtain the
following relations within the first order approximation:
 \begin{eqnarray}
 \eta&=&b\epsilon_{1}-\frac{a-(d+c)}{\sqrt{2}}\epsilon_{2}, \\
 \xi&=&-2d\epsilon_{1}-\sqrt{2}b\epsilon_{2}.
\end{eqnarray}
Judging from the range of $\epsilon_{1}$ and $\epsilon_{2}$, we find
that $\eta$ and $\xi$ are much smaller than the elements of ${\cal
M}_{\nu}^{0}$. We therefore assume that $\eta$ and $\xi$ are zero in the
present paper.  We further assume $c=0$ for simplicity.  In this case,
the structure of the mass matrix belongs to the type-C texture with two
independent zeros, defined in Ref. \cite{Fra3}, and the phenomenology
is discussed in Refs. \cite{Xin1,Xin2,Des}.  We obtain the three mass
eigenvalues and $\epsilon_{3}$ as follows:
\begin{eqnarray}
   \hat{M_{\nu}} \simeq U^{T}_{0}{\cal M}_{\nu}^{0}U_{0}
   &=& \left(\begin{array}{ccc}
    \frac{a-d+\sqrt{(a+d)^2+8b^2}}{2}  & 0    & 0    \\
      0        &   \frac{a-d-\sqrt{(a+d)^2+8b^2}}{2}  &  0     \\ 
   0   &   0   &  d
	\end{array}\right)
	\equiv\left(\begin{array}{ccc}
    m_{1}  & 0    & 0    \\
      0        &   m_{2}  &  0     \\ 
   0   &   0   &  m_{3}
	\end{array}\right), \\
	\mbox{tan2}\epsilon_{3}&=&-\frac{a+d}{2\sqrt{2}b}. \label{tan}
\end{eqnarray}
Since we identify $\Delta_{a}$ and $\Delta_{s}$ as
$\Delta_{a} \equiv |m_{1}^{2}-m_{3}^{2}| \simeq |m_{2}^{2}-m_{3}^{2}|$ and $\Delta_{s} \equiv |m_{1}^{2}-m_{2}^{2}|$,
we obtain the ratio $R$,
\begin{eqnarray}
R=\frac{\Delta_{s}}{\Delta_{a}}=\frac{4\Bigl|(a-d)\sqrt{(a+d)^2+8b^2}\Bigr|}
{\Bigl|\Bigl[a-d+\sqrt{(a+d)^2+8b^2}\Bigr]^2-4d^2\Bigr|}. \label{R}
\end{eqnarray}
\par It is shown in Ref. \cite{Fra2} that $a$ must have an appropriate
magnitude obtained from the experimental data mentioned above.  This can be
understood as follows.  If we set $a=0$, we obtain a simple relation
between $\tan2\epsilon_{3}$ and R,
\begin{eqnarray}
\tan2\epsilon_{3}=\frac{1}{2\sqrt{2}}\biggl|\frac{d}{b}\biggr|\simeq \frac{1}{4}R, \label{p=0}
 \end{eqnarray}
where $b \simeq -\sqrt{\Delta_{a}/2}$ and $d\simeq \frac{1}{2}\sqrt{\Delta_{a}}R$.
On the other hand, R and $\tan 2\epsilon_{3}$ must satisfy 
\begin{eqnarray}
3.2 \times 10^{-3} < \frac{1}{4}R < 3.0 \times 10^{-2},\  \ 7.7 \times 10^{-2} < \tan 2\epsilon_{3} < 0.66
\label{p=1} 
\end{eqnarray}
from the data (\ref{R3}).  We therefore find that $a$ must have an
appropriate magnitude because the conditions (\ref{p=0}) and
(\ref{p=1}) are incompatible.  Similarly, we can show that $d$ must also
have an appropriate magnitude.  From Eqs. (\ref{tan}), (\ref{R}),
and (\ref{R3}), we find that mild fine-tuning $|a+d| \gg |a-d|$
is needed to fit the model with the LMA MSW solution.

Because the difference $a-d$ $(\equiv \epsilon)$ is small in comparison
with $b,a,d,$ it can be treated as a small perturbation.  We obtain the
following values for $b,a,d,$ and $\epsilon$:
\begin{eqnarray}
 b &\simeq& -\sqrt{\frac{\Delta_{a}}{2}} \simeq -3.5 \times 10^{-2} \ \mbox{eV}, \label{a}\\
 a &\simeq& (\tan 2 \epsilon_{3}) \sqrt{\Delta_{a}} \simeq 2.0 \times 10^{-2}  \ \mbox{eV}, \label{p}\\
 d &\simeq& (\tan 2 \epsilon_{3}) \sqrt{\Delta_{a}}-\epsilon \simeq 1.9 \times 10^{-2} \ \mbox{eV}, \label{c}\\
 \epsilon  &\simeq& \frac{1}{2}\sqrt{\Delta_{a}}R\frac{1}{\sqrt{1+(\tan 2 \epsilon_{3})^2}} 
 \simeq 6.5 \times 10^{-4} \ \mbox{eV}, 
 \end{eqnarray}
for the best fit values $\Delta_{a}=2.5 \times 10^{-3} \ \mbox{eV}^{2},\ \epsilon_{3}=0.19$, and $R=2.8 \times 10^{-2}$.

\subsection{GZM neutrino mass matrix  \label{B}}
In the present section, we review the GZM \cite{Zee,Gri}.  The interactions of the GZM are the following:
 \begin{eqnarray}
 {\cal L}_{SM}^{Yukawa}&=& \sum\limits_{i=1,2} \vec{\bar{l}}_{L,a}
 \Gamma_{i} \Phi_{i,a} \vec{e}_{R} +\mbox{H.c.}, \label{sm}\\ 
 {\cal L}_{Zee}^{Yukawa} &=&\vec{\bar{l}}_{L,a} i(\sigma_{2})_{ab}f
 \vec{l}_{L,b}^{c} h^{-} +\mbox{H.c.}, \\
 {\cal L}_{Higgs}^{cubic}&=&\lambda \Phi_{1,a}^{T} i(\sigma_{2})_{ab}
 \Phi_{2,b}h^{-} +\mbox{H.c.} \label{higgs} 
 \end{eqnarray} 
 Here, $h^{-}$ is the Zee singlet, $\vec{l}_{L}$ is the lepton doublet in a
 weak eigenstate, $\Phi_{1}$ and $\Phi_{2}$ are the two Higgs
 doublets, $\Gamma_{1}$ and $\Gamma_{2}$ are two Yukawa coupling
 matrices, $f$ is an antisymmetric coupling matrix, $\lambda$ is a
 cubic coupling constant, and $a,b$ are indices of the $SU(2)_{L}$
 doublets.  \par We write the vacuum expectation value (VEV) of each
 Higgs doublet as $\langle\phi_{i}^{0}\rangle=v_{i}/\sqrt{2}$ ($i$ = 1,2).
 Rotating $\Phi_{1}$ and $\Phi_{2}$ as follows:
\begin{eqnarray}
\left(\begin{array}{r}
                                      \Phi_{1}^{'} \\
                                      \Phi_{2}^{'}           
                                   	  \end{array}\right)
									   = \left(\begin{array}{cc}
                                 \cos \beta & -\sin \beta \\
                                  \sin \beta & \cos \beta 
                                   	\end{array}\right)
									   \left(\begin{array}{r}
                                 \Phi_{1} \\
                                      \Phi_{2} 
                                   	\end{array}\right),   \label{beta}
									    \end{eqnarray}
where $\tan\beta=v_{1}/v_{2}$, we make only one of the two Higgs
 doublets have the VEV, i.e.,  $\langle{\phi^{'}}_{1}^{0}\rangle=0$ and
 $\langle{\phi^{'}}_{2}^{0}\rangle=v/\sqrt{2}=\sqrt{(v_{1}^{2}+v_{2}^{2})/2}$.  While
 $\phi_{2}^{'\pm}$ is the would-be Nambu-Goldstone boson, the
 diagonalization of the mass matrix for the remaining physical charged
 Higgs fields $h^{+}$ and $\phi_{1}^{'+}$ goes as
 \begin{eqnarray}
 {\cal L}_{Higgs} \supset \left(\begin{array}{rr}
                                     h^{+} ,& \phi_{1}^{'+}
									 \end{array}\right)
									 \left(\begin{array}{rr}
                        M_{h}^{2}     &  -\frac{\lambda v}{\sqrt{2}}  \\
                         -\frac{\lambda v}{\sqrt{2}}           &   (M_{1}^{'+})^{2}  
						\end{array}\right) 
						\left(\begin{array}{r}
                                     h^{-} \\
									 \phi_{1}^{'-}
									 \end{array}\right)
=\left(\begin{array}{rr}
                                     S_{1}^{+} ,& S_{2}^{+}
									 \end{array}\right)
									 \left(\begin{array}{rr}
                        M_{1}^{2}     &  0  \\
                         0           &    M_{2}^{2} 
						\end{array}\right) 
						\left(\begin{array}{r}
                                     S_{1}^{-} \\
									 S_{2}^{-}
									 \end{array}\right),
 \end{eqnarray}
 using the orthogonal transformation
 \begin{eqnarray}
\left(\begin{array}{r}
                                      S_{1}^{-} \\
                                      S_{2}^{-}           
                                   	  \end{array}\right)
									   = \left(\begin{array}{cc}
                                 \cos \alpha & -\sin \alpha \\
                                  \sin \alpha & \cos \alpha 
                                   	\end{array}\right)
									   \left(\begin{array}{r}
                                h^{-} \\
                                      \phi_{1}^{'-} 
                                   	\end{array}\right),   \label{alpha}
 \end{eqnarray}
where the rotational angle $\alpha$ and the two mass eigenvalues $M_{1},M_{2}$ are determined as 
\begin{eqnarray}
\tan 2\alpha &=& \frac{-\sqrt{2}\lambda v}{(M_{1}^{'+})^{2}-M_{h}^{2}}, \\
(M_{1,2})^2 &=& \frac{M_{h}^{2}+(M_{1}^{'+})^{2}\pm \sqrt{[M_{h}^{2}-(M_{1}^{'+})^2]^2+2\lambda^2v^2}}{2}.
 \end{eqnarray}
 Using $\Phi_{1}^{'}$ and $\Phi_{2}^{'}$ in Eq. (\ref{beta}), the Lagrangian density (\ref{sm}) is written as 
  \begin{eqnarray}
 {\cal L}_{SM}^{Yukawa}&=& \vec{\bar{l}}_{L,a} \Bigl[\big(\cos\beta\ \Gamma_{1}
  -\sin \beta\ \Gamma_{2} \bigr) \Phi_{1,a}^{'} + \bigl(\sin \beta\ \Gamma_{1}
 +\cos \beta\ \Gamma_{2}  \bigr) \Phi_{2,a}^{'} \Bigr]\vec{e}_{R} +\mbox{H.c.} 
 \end{eqnarray}
 Since only $\phi_{2}^{'0}$ has the VEV, using the biunitary transformation
 \begin{eqnarray}
 \vec{l}_{L,a}^{m}&\equiv& V_{L}\vec{l}_{L,a},\ 
  \vec{l}_{L,1}^{m}\equiv  \vec{\nu}_{L}= \left(\begin{array}{c}
                                          \nu_{e}\\   
       \nu_{\mu}  \\ 
         \nu_{\tau}
                                   	  \end{array}\right)_L, \ 
										  \vec{l}_{L,2}^{m}\equiv \vec{e}_{L}^{m}= \left(\begin{array}{c}
                                               e \\   
       \mu \\ 
        \tau
                                   	  \end{array}\right)_L,\\
  \vec{e}_{R}^{m}&\equiv& V_{R}\vec{e}_{R}=\left(\begin{array}{c}
    	e\\   
       \mu  \\ 
         \tau
	\end{array}\right)_{R}.
 \end{eqnarray}
The matrix $(\sin \beta\ \Gamma_{1}+\cos \beta\ \Gamma_{2})$ is diagonalized as
  \begin{eqnarray}
 \hat{\Gamma}^{m}\equiv V_{L}\bigl(\sin \beta\ \Gamma_{1}+\cos \beta\ \Gamma_{2} \bigr)V_{R}^{\dagger} 
 \equiv \frac{\sqrt{2}}{v}\hat{M}_{l-m}=\frac{\sqrt{2}}{v}\left(\begin{array}{ccc}
                                    m_{e} &0        & 0\\
                                     0    & m_{\mu} & 0\\
									 0    &0        & m_{\tau}
									 \end{array}\right),
 \end{eqnarray} 
 where $\hat{M}_{l-m}$ is the mass matrix of the charged leptons. 
 The Lagrangian density (\ref{sm}) is written as
  \begin{eqnarray}
 {\cal L}_{SM}^{Yukawa}&=& \vec{\bar{l}}_{L,a}^{m} \Bigl[\big( \cos \beta\ \Gamma_{1}^{m}
  -\sin \beta\ \Gamma_{2}^{m} \bigr) \Phi_{1,a}^{'} + \hat{\Gamma}^{m} \Phi_{2,a}^{'} \Bigr]\vec{e}_{R}^{m} +
  \mbox{H.c.},
  \end{eqnarray} 
where $\Gamma_{i}^{m} \equiv V_{L} \Gamma_{i} V_{R}^{\dagger}$ ($i$ = 1,2). 
\par
Since we estimate the interaction rates in the early Universe in Sec. \ref{Cos},
 we write down the Lagrangian density in the
 symmetric phase ($v=0$) in addition to that in the symmetry broken phase ($v=246 $GeV) 
 where the neutrinos have the Majorana masses.
\begin{itemize}
 \item[1.] Symmetric phase ($v=0$) \\ The interaction Lagrangian density
 is as follows:
 \begin{eqnarray} {\cal L}_{SM}^{Yukawa}&=&
 \vec{\bar{l}}_{L,a}^{m} \Bigl[\big( \frac{1}{\tan \beta}
 \hat{\Gamma}^{m} - \frac{1}{\sin \beta}\Gamma_{2}^{m} \bigr)
 \Phi_{1,a}^{'} + \hat{\Gamma}^{m}\Phi_{2,a}^{'} \Bigr]
 \vec{e}_{R}^{m}+\mbox{H.c.}, \\ {\cal L}_{Zee}^{Yukawa}
 &=&2\vec{\bar{\nu}}_{L}^{m} f^{m} (\vec{e}_{L}^{m})^{c} h^{-}
 +\mbox{H.c.}, \\ {\cal L}_{Higgs}^{cubic}&=&\lambda
 {\Phi^{'}}_{1,a}^{T} i(\sigma_{2})_{ab} {\Phi^{'}}_{2,b}h^{-}
 +\mbox{H.c.}  \end{eqnarray} Here, all leptons are massless and
 $f^{m} \ (\equiv V_{L}fV_{L}^{T})$ is antisymmetric.
 
 \item[2.] Symmetry broken phase($v=246$ GeV) \\
 The interaction Lagrangian density is as follows:
 \begin{eqnarray}
 {\cal L}_{SM}^{Yukawa}&=& \vec{\bar{\nu}}_{L}^{e} \Bigl[\bigl(
  \frac{1}{\tan \beta} \hat{\Gamma}^{m} - \frac{1}{\sin \beta}
  \Gamma_{2}^{m} \bigr) (-\sin \alpha S_1^{+}+\cos \alpha S_2^{+})+
  \hat{\Gamma}^{m} \phi_{2}^{'+} \Bigr] \vec{e}_{R}^{m} \nonumber \\
  &&+ \vec{\bar{e}}_{L}^{m} \Bigl[\bigl( \frac{1}{\tan \beta}
  \hat{\Gamma}^{m} - \frac{1}{\sin \beta} \Gamma_{2}^{m} \bigr)
  \phi_{1}^{'0} +\hat{\Gamma}^{m}
  \bigl(\frac{v+\sigma+iN_{2}}{\sqrt{2}}\bigr) \Bigr]\vec{e}_{R}^{m}
  +\mbox{H.c.}, \\ {\cal L}_{Zee}^{Yukawa} &=&2\vec{\bar{\nu}}_{L}^{m}
  f^{m} (\vec{e}_{L}^{m})^{c} (\cos \alpha S_1^{-} +\sin \alpha
  S_{2}^{-})+\mbox{H.c.}, \\ {\cal L}_{Higgs}^{cubic}&=&\lambda (\cos
  \alpha S_1^{-}+\sin \alpha S_{2}^{-}) \Bigl[
  \frac{1}{\sqrt{2}}(\sigma +iN_{2})(-\sin \alpha S_1^{+}+\cos \alpha
  S_{2}^{+}) \nonumber \\ &&-\phi_{1}^{'0}\phi_{2}^{'+}
  \Bigr]+\mbox{H.c.}  \end{eqnarray} 
Here, the Higgs fields
  $S_{1}^{-},S_{2}^{-}$ have masses $M_{1},M_{2}$, respectively.  The fields
  $\phi_{2}^{'+}$ and $N_{2}$ are the would-be Nambu-Goldstone bosons.
  \end{itemize} \par
\begin{figure}[h]
\begin{center}
\unitlength=1mm
\begin{picture}(230,50)
\epsfile{file=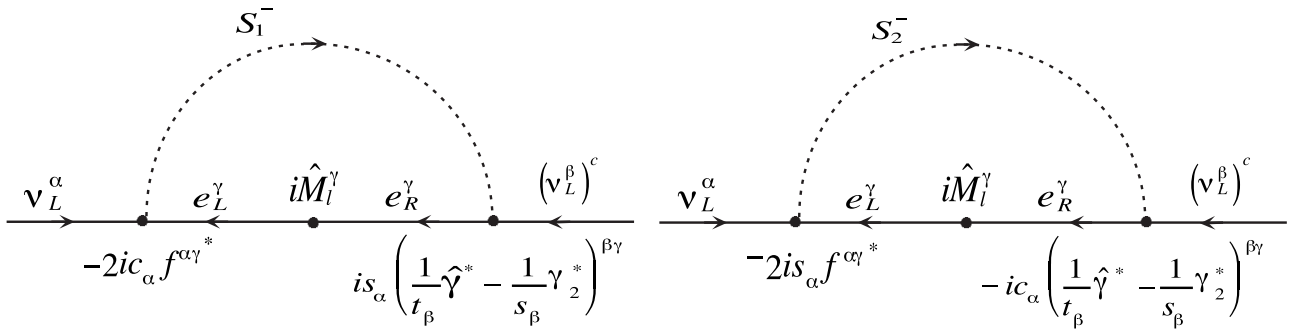,height=4.3cm}
\end{picture}
\caption{The diagrams that are used to calculate the radiatively
induced neutrino mass matrix. \label{majorana}
We use $s_{\alpha}\equiv \sin \alpha, c_{\alpha}\equiv \cos \alpha$,
and $t_{\alpha}\equiv \tan \alpha$ and so on in this figure.}
\end{center}
\end{figure}

 In this phase, the Majorana neutrino mass is
  calculated from the two one-loop diagrams in
  Fig. \ref{majorana} and the transposed diagrams of them, for
  $m_{e},m_{\mu},m_{\tau} \ll M_{1},M_{2}$,
\begin{eqnarray}
 i{\cal M}^{\alpha \beta}&=&A\Bigl[ (f^{*} \hat{M_{l}^{2}}+\hat{M_{l}^{2}} f^{\dagger} ) -\frac{v}{\sqrt{2} \cos \beta}
 (f^{*} \hat{M_{l}}\ {\Gamma_{2}^{m}}^{\dagger} +\Gamma_{2}^{m*}\hat{M_{l}}\ f^{\dagger} ) 
 \Bigr]^{\alpha \beta}  \label{gzm}  \\
(&=& i{{\cal M}^{T}}^{\alpha \beta}), \nonumber
 \end{eqnarray}
where 
 $A \equiv (1/8\sqrt{2}\pi^{2} v \tan \beta) \sin 2 \alpha \log(M_{2}^{2}/M_{1}^{2})$
  and $\gamma_{2}^{\alpha \beta}, \hat{\gamma}^{\alpha \beta}$ are elements of the matrices
   $\Gamma_{2}^{m},\hat{\Gamma}^{m}$, respectively.

\subsection{Matching the GZM with the experiments \label{match}}
We obtained the relation between the value of $a$ and the experimental
data in Sec. \ref{A},
\begin{eqnarray}
  a=\tan 2\epsilon_{3} \sqrt{\Delta_{a}}. \label{ex}
\end{eqnarray}
On the other hand, we calculated the radiatively induced $a$ in the GZM in Sec. \ref{B},
\begin{eqnarray}
  a=-2AB[m_{\mu}{f^{*}}^{e \mu}{{\gamma}^{*}}_{2}^{e \mu}+m_{\tau}{f^{*}}^{e \tau}{{\gamma}^{*}}_{2}^{e \tau}], \label{th}
\end{eqnarray}
where $B=v/\sqrt{2}\cos \beta$. From Eq. (\ref{ex}) and
Eq. (\ref{th}), we find that one of ${\gamma}_{2}^{e \tau}$ and
${\gamma}_{2}^{e \mu}$ must be nonzero at least.  Because we
prefer a minimal extension here, we assume that the other elements of
$\Gamma_{2}^{m}$ are zero.  From now on, we further assume that $f$ and
$\Gamma_{2}^{m}$ are real.  Here, we match the neutrino mass matrix in
the GZM (\ref{gzm}) with that determined from the experimental data
(\ref{a})$-$(\ref{c}) as follows
   \begin{eqnarray}
  b &=&-\sqrt{\frac{\Delta_{a}}{2}}=A[f^{e \mu} m_{\mu}^{2}-Bm_{\tau}f^{\mu \tau}  {\gamma}_{2}^{e \tau} ]
  =-A[f^{e \tau} m_{\tau}^{2} +Bm_{\mu}f^{\mu \tau}  {\gamma}_{2}^{e \mu} ], \label{m1} \\
  a&=& \tan 2\epsilon_{3}\sqrt{\Delta_{a}}=-2AB[m_{\mu}f^{e \mu}{\gamma}_{2}^{e \mu}  + m_{\tau}f^{e \tau}{\gamma}_{2}^{e \tau} ] \label{m2}, \\
  d&=& \tan 2\epsilon_{3}\sqrt{\Delta_{a}}-\epsilon= Af^{\mu \tau}m_{\tau}^{2} \label{m3}.
\end{eqnarray}
In this model, there are nine parameters $(M_{1}, M_{2}, \alpha, \beta, f^{e \mu}, f^{e \tau}, 
f^{\mu \tau}, {\gamma}_{2}^{e \tau}, {\gamma}_{2}^{e \mu} )$. 
From Eqs. (\ref{m1}) $-$ (\ref{m3}), we parametrize four parameters 
 $(f^{e \mu},f^{e \tau}, f^{\mu \tau},{\gamma}^{e \tau}_{2})$ using the other five parameters
  $(M_{1}, M_{2}, \alpha, \beta,{\gamma}_{2}^{e \mu} )$ as follows
\begin{eqnarray}
  f^{e \mu} &=&\frac{1}{A} \frac{b\bigl[b + B d (m_{\mu}/m_{\tau}^2)  {\gamma}_{2}^{e \mu}\bigr] + \frac{1}{2}da}  
  {m_{\mu}^{2} \bigl[b + Bd (m_{\mu}/m_{\tau}^{2}){\gamma}_{2}^{e \mu}\bigr]
  - Bd m_{\mu} {\gamma}_{2}^{e \mu}}, \label{f1} \\
  f^{e \tau} &=& -\frac{b}{A m_{\tau}^2} - \frac{B}{A}d\frac{m_{\mu}}{m_{\tau}^{4}}{\gamma}_{2}^{e \mu}, \label{f2}\\
  f^{\mu \tau} &=& \frac{d}{A m_{\tau}^2},  \label{f3} \\
  {\gamma}^{e \tau}_{2} &=& \frac{m_{\tau}}{m_{\mu}} \frac{b{\gamma}_{2}^{e \mu}+ (a/2B)m_{\mu}}{b +
    Bd{\gamma}_{2}^{e \mu} (1/m_{\tau}^{2}-1/m_{\mu}^{2}) \label{g1}
	}.
\end{eqnarray}
We consider the simplest two cases, (1)$ {\gamma}^{e \tau}_{2} \not=0
 ,{\gamma}^{e \mu}_{2}=0$ and (2) ${\gamma}^{e \tau}_{2}=0,{\gamma}^{e
 \mu}_{2} \not=0$.  In these cases, we have the following relations.
\begin{itemize}
  \item[(1)]  ${\gamma}^{e \tau}_{2} \not=0 ,{\gamma}^{e \mu}_{2}=0$.
  Equations (\ref{f1})$-$(\ref{g1}) are written as
  \begin{eqnarray}
  f^{e \mu}&=&\frac{b}{Am_{\mu}^2}+\frac{da}{2Abm_{\mu}^2} 
  \simeq -3.7\times 10^{-5}\Biggl(\frac{10^{-4} \ \mbox{GeV}^{-1}}{A}\Biggr), \nonumber 
  \\
  f^{e \tau}&=&-\frac{b}{A m_{\tau}^2}\simeq 1.1\times 10^{-7}
  \Biggl(\frac{10^{-4} \ \mbox{GeV}^{-1}}{A}\Biggr), \label{gm1} \\
  f^{\mu \tau}&=&\frac{d}{A m_{\tau}^2}\simeq 6.1\times 10^{-8}
  \Biggl(\frac{10^{-4} \ \mbox{GeV}^{-1}}{A}\Biggr),\ 
  {\gamma}^{e \tau}_{2}=\frac{am_{\tau}}{2Bb}\simeq -2.1\times 10^{-3}\Biggl(\frac{\cos \beta}{1/\sqrt{2}}\Biggr). 
  \nonumber  
\end{eqnarray}
\item[(2)] ${\gamma}^{e \tau}_{2}=0, \ {\gamma}^{e \mu}_{2} \not=0$. Equations
 (\ref{f1})$-$(\ref{g1}) are written as 
\begin{eqnarray}
  f^{e \mu}&=&\frac{b}{Am_{\mu}^2}\simeq -3.2\times 10^{-5}\Biggl(\frac{10^{-4} \ 
  \mbox{GeV}^{-1}}{A}\Biggr), \nonumber \\
  f^{e \tau}&=&-\frac{b}{A m_{\tau}^2}+\frac{m_{\mu}^2}{m_{\tau}^4}\frac{da}{2Ab}
  \simeq -\frac{b}{A m_{\tau}^2} \simeq 1.1\times 10^{-7}\Biggl(\frac{10^{-4} \ 
  \mbox{GeV}^{-1}}{A}\Biggr), \label{gm2}\\
  \ f^{\mu \tau}&=&\frac{d}{A m_{\tau}^2}\simeq 6.1 \times 10^{-8}\Biggl(
  \frac{10^{-4} \ \mbox{GeV}^{-1}}{A}\Biggr),\ 
  {\gamma}^{e \mu}_{2}=-\frac{am_{\mu}}{2Bb}\simeq 1.2 \times 10^{-4}
  \Biggl(\frac{\cos \beta}{1/\sqrt{2}}\Biggr).  
 \nonumber  
\end{eqnarray}
\end{itemize}

\subsection{Experimental constraints on $\tau(\mu)$ decay, 
$\tau(\mu)$ $\rightarrow$ e $\gamma$ and 0$\nu\beta \beta$
 amplitude} 

In the present section, we examine whether the parameters obtained in
the previous section [Eqs. (\ref{gm1}),(\ref{gm2})] are consistent with
the constraints from the processes $\tau(\mu)$ decay, $\tau(\mu)$
$\rightarrow$ e$\gamma$, and 0$\nu\beta\beta$ decay \cite{Gri}.
\begin{itemize}
\item[1.] $\tau(\mu)$ decay \par
 Since the leptonic flavor changing neutral current does not exist in the standard model at all, 
 stringent constraints are imposed on their decay amplitudes \cite{Particle}.
 There is, however, $\tau(\mu)$ decay through the FCNCs at the tree level in the GZM.
 These constraints are satisfied for the parameter set (\ref{gm1}) and (\ref{gm2}) as follows
 for $\beta \simeq \pi/4$:
\begin{eqnarray}
\mbox{Br}(\tau^{-} \rightarrow e^{-}\mu^{-} \mu^{+}) \simeq \frac{|{\gamma}_{2}^{e\tau}
 (\sqrt{2}m_{\mu}/v)|^2}{G_{F}^{2} ({M^{'}}_{1}^{0})^{4}}&\lesssim& 1.1 \times 10^{-10} \nonumber\\
 &\ll& 1.8 \times 10^{-6} \ (\mbox{experiment \cite{Particle}}), \nonumber\\
 \mbox{Br}(\mu^{-} \rightarrow e^{-}e^{-}\mu^{+}) \simeq \frac{|{\gamma}_{2}^{e\mu}
 (\sqrt{2}m_{e}/v)|^2}{G_{F}^{2} ({M^{'}}_{1}^{0})^{4}}
 &\lesssim& 8.7 \times 10^{-18}  \nonumber\\
 &\ll& 1.0 \times 10^{-12} \ (\mbox{experiment \cite{Particle}}),
 \end{eqnarray}
where $G_{F}$ is the Fermi coupling constant and ${M^{'}}_{1}^{0}\gtrsim 100 \ \mbox{GeV}$.
 Since the coupling constants $f^{e \mu}, f^{e \tau}$, and $f^{\mu \tau}$ are much smaller than 
 $\gamma_{2}^{e\tau}$ and $\gamma_{2}^{e\mu}$, the contributions to the branching ratios purely from $f$ are negligible \cite{Pet,Smi,Mit}.
 
\item[2.] $\tau(\mu)$ $\rightarrow$ e $\gamma$ \par
Since these processes violate the lepton flavor, they do not exist in the standard model at all.
 On the other hand, these processes become possible in the GZM 
 through the exchange of the charged and neutral Higgs bosons. 
 Since they are rare events, stringent constraints are imposed on the amplitudes \cite{Particle}.
  These constraints are safely satisfied for the parameter set (\ref{gm1}) and (\ref{gm2}) as follows 
  for $\beta \simeq \pi/4$:
  \begin{eqnarray}
  \mbox{Br}(\tau^{-} \rightarrow e^{-} \gamma) &\simeq& \alpha \frac{|{\gamma}_{2}^{e\tau}
  (\sqrt{2}m_{\tau}/v)|^2}
  {G_{F}^{2} ({M^{'}}_{1}^{0})^{4}}
  \lesssim 2.3 \times 10^{-10}  \ll 2.7 \times 10^{-6} \ (\mbox{experiment \cite{Particle}}),  \nonumber \\
  \mbox{Br}(\mu^{-} \rightarrow e^{-} \gamma) &\simeq& \alpha \frac{|{\gamma}_{2}^{e\mu}
 (\sqrt{2}m_{\mu}/v)|^2}
  {G_{F}^{2} ({M^{'}}_{1}^{0})^{4}}
  \lesssim 2.8 \times 10^{-15}  \ll 1.2 \times 10^{-11} \ (\mbox{experiment \cite{Particle}}). \nonumber \\
\end{eqnarray} 
 Again, since the coupling constants $f^{e \mu}, f^{e \tau}$, and $f^{\mu \tau}$ are much smaller than
 $\gamma_{2}^{e\tau}$ and $\gamma_{2}^{e\mu}$, the contributions to the branching ratios purely
 from $f$ are negligible \cite{Pet,Smi,Mit}.

\item[3.] 0$\nu\beta \beta$ \\ The observation of the neutrinoless
double beta decay (0$\nu\beta\beta$) gives evidence that the
neutrinos have Majorana masses. The study of this process is 
a crucial test of the validity of our model, since neutrinos
have Majorana masses, the interaction of the Zee singlet $h^{-}$
violates lepton number, and the Yukawa coupling $\Gamma_{2}^{m}$ is
flavor changing. As was discussed by Schechter and
Valle \cite{Valle-beta} in the theory with an exotic doubly charged
scalar, the extension of the Higgs sector possibly provides a new
source of the decay.  We can, in fact, think of a process with the
exchange of our exotic scalar, i.e., $h^{-}$ [see Fig. \ref{betafig}(a)].
However, we readily know that this diagram does not exist in our
 model. This is because the Yukawa coupling of $h^{-}$ is flavor
off-diagonal and the intermediate neutrinos are either $\nu_{\mu}$ or
$\nu_{\tau}$.  This, in turn, means that the Yukawa coupling of $\phi_{1}^{'-}$ at
another vertex should be flavor changing, i.e., ${\gamma}^{\mu
e}_{2}$ or ${\gamma}^{\tau e}_{2}$, which vanishes
in our model.  Another source of the double beta decay comes from the
exchange of Majorana neutrinos with Majorana mass insertion due to
$W^{\pm}$ exchange. As was first pointed out by
Wolfenstein \cite{Wol2}, the amplitude is proportional to $m_{ee}$. The
absence of any report of this process so far gives the upper bounds to the
element $m_{ee}$ in the neutrino mass matrix
\cite{Particle} \footnote{It was recently reported that the first
evidence for 0$\nu\beta \beta$ has been observed \cite{0}. The value is
$|m_{ee}|=0.11 \sim 0.56$ eV (95\% C.L.) with the best fit value
0.39 eV.  There are also, however, arguments against this report \cite{Aal}.},
\begin{eqnarray}
  |m_{ee}| < 0.1 \ \mbox{eV}  \ (90 \% \ \mbox{C.L.}). \label{mee}
\end{eqnarray}
The value obtained in Sec. \ref{A} is consistent with this upper
bound as follows:
\begin{eqnarray}
  m_{ee}=a \simeq 2.0 \times 10^{-2} \ \mbox{eV}.
\end{eqnarray}
Since the Yukawa coupling $\Gamma_{2}^{m}$ is flavor changing, another
source of decay exists as shown in Fig. \ref{betafig}(b), where 
$m_{\mu \mu}$ or $m_{\tau \tau}$ contributes, but not $m_{ee}$. This
diagram, however, does not exist from reasoning similar to the above;
namely, the amplitude of the diagram is also proportional to
${\gamma}^{\mu e}_{2}$ or ${\gamma}^{\tau e}_{2}$,
 which does not exist.  Thus the ordinary process due to $m_{ee}$
discussed above is the unique source of double beta decay. If the
sensitivity of the experiment is improved by a factor of 5 or so, the decay
should be seen if our model is valid. 

\begin{figure}[h]
\begin{center}
\unitlength=1mm
\begin{picture}(130,60)
\epsfile{file=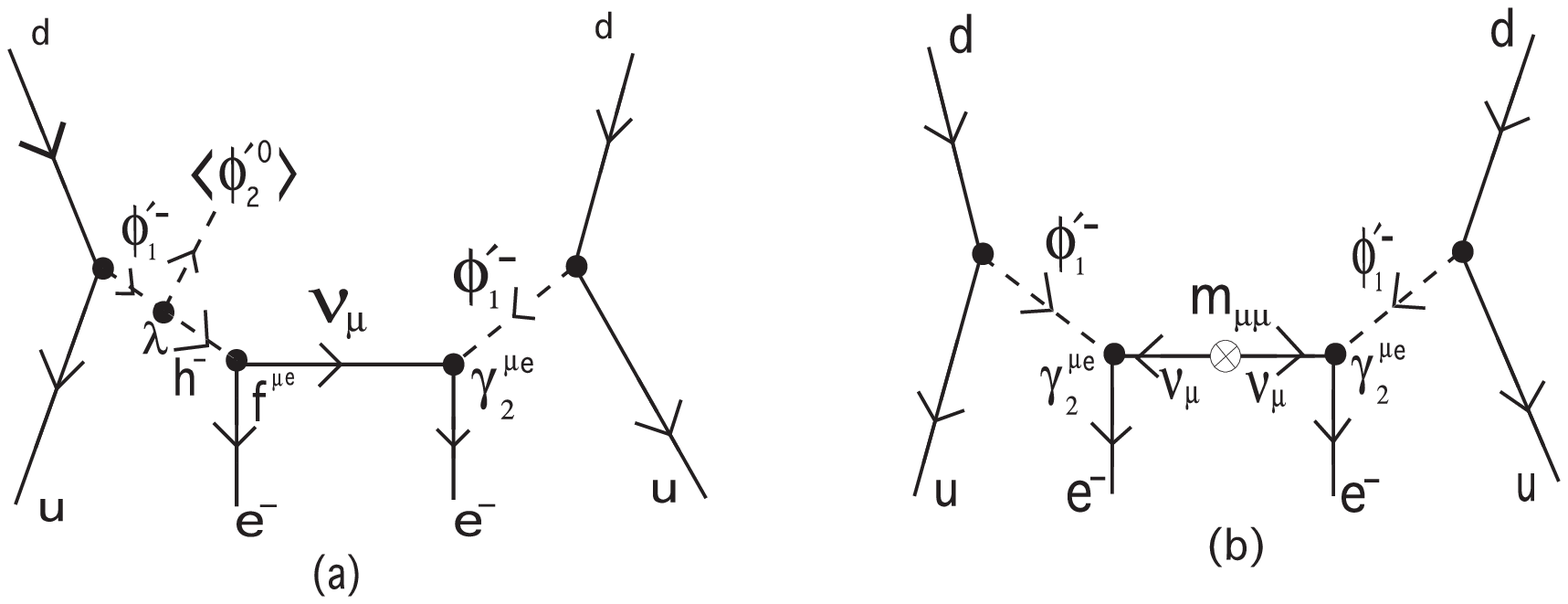,height=5.0cm}
\end{picture}
\caption{The diagrams contributing to neutrinoless
double beta decay. In (a), the $h^{-}$ interaction violates
 the lepton number, while in (b) the lepton number is violated by the insertion of
 $m_{\mu \mu}$. Both processes fail to exist because of the
vanishing Yukawa coupling ${\gamma}^{\mu e}_{2}=0$.
In both diagrams $\nu_{\mu}$ may be replaced by $\nu_{\tau}$. \label{betafig}}
\end{center}
\end{figure}

\end{itemize}

\section{Cosmological implications of the GZM \label{Cos}}
In Sec. \ref{II}, we obtained the values for the coupling constants
($f^{e \mu},f^{e \tau}, f^{\mu \tau},{\gamma}^{e \tau}_{2},{\gamma}^{e
\mu}_{2}$) suggested by the neutrino oscillation data.  In the present
section, we investigate whether the primordial baryon number is washed
out or not for these values.  If there are some leptonic conserved
numbers, the primordial baryon number remains finite even though the
sphaleron process partially washes out the baryon number.  We therefore
 consider what leptonic numbers are effectively conserved at
a high temperature.  We then calculate what amount of the primordial
baryon number remains in the case that there exist some conserved
numbers.  We also briefly consider the constraints from the recent CMB
observation \cite{Wmap1,Wmap2,Wmap3,Wmap4} and the possibility of the
Z-burst scenario in this model \cite{Wei,Far}.

\subsection{Conserved number for each coupling}
It is convenient to define the following global $U(1)$ charges as
\begin{eqnarray}
L_{e \mu} \equiv L_{\tau}-L_{e}-L_{\mu} ,\  \ L_{e \tau} \equiv L_{\mu}-L_{e}-L_{\tau} ,\ \
   L_{\mu \tau} \equiv L_{e}-L_{\mu}-L_{\tau}
\end{eqnarray}
for an investigation into the washing out of the baryon number.
These quantities are conserved numbers in the standard model. We,
however, have coupling constants which violate these conserved
numbers in our model.  Each coupling conserves two of the three
independent numbers which are shown in the Table \ref{conserve} using
a linear combination of $L_{e\mu},L_{e\tau}$, and $L_{\mu \tau}$.

\begin{table}[h] \begin{tabular}{|c|c|c|c|c|c|}\hline
Coupling constant & $f^{e \mu}$ & $f^{e \tau}$ & $f^{\mu \tau}$ &
${\gamma}^{e \tau}_{2}$ & ${\gamma}^{e \mu}_{2}$ \\ \hline 
Conserved number & $L_{e \tau}$, $L_{\mu \tau}$ & $L_{e \mu}$, $L_{\mu
\tau}$ & $L_{e \mu}$, $L_{e \tau}$ & $L_{e \tau}$, $L_{e \mu}+L_{\mu \tau}$ &
$L_{e \mu}$, $L_{e \tau}+L_{\mu \tau}$ \\ \hline 
\end{tabular}
\caption{Conserved  number for each coupling constant. \label{conserve}}
\end{table}

\subsection{Interaction rates induced by each coupling}
 We here estimate the interaction rates induced by the coupling constants $f$ and $\gamma_{2}$. 

\begin{center}1. $L_{\mu \tau}$ violating process induced by the coupling constant $f^{\mu \tau}$ 
\end{center}
\par
We consider the $h^{-}$ decay process and the inverse decay process induced by
 the coupling constant $f^{\mu \tau}$.
The interaction rates for these processes are estimated at a temperature
      T $\gtrsim M_{h}$ \cite{Ear,Haba},
\begin{eqnarray}
\Gamma(h^{-} \leftrightarrow \mu^{-} \nu_{\tau}) &=& \frac{1}{2\pi}|f^{\mu \tau}|^{2}\frac{M_{h}^{2}}{E} \simeq
 \frac{1}{5.4\pi}|f^{\mu \tau}|^{2}\frac{M_{h}^{2}}{T},  \label{hdecay}
\end{eqnarray}
where $M_{h}$ is the mass of the Zee singlet and  $\langle E \rangle \simeq$ 2.7 T (boson)
 is the averaged energy.
Similar results are obtained for the processes induced by the coupling constants $f^{e \mu}$ and $f^{e \tau}$.

\begin{center} 2. Processes induced by the coupling constant ${\gamma}_{2}^{e\tau}$ \end{center}
\par
 We next estimate the interaction rates for the processes
induced by the coupling constant ${\gamma}_{2}^{e \tau}$.  We
estimate the interaction rates of ${\phi^{'}}_{1}^{-}$ decay and the
inverse decay.  The interaction rate for the process
${\phi^{'}}_{1}^{-}\leftrightarrow(\nu^{e}_{L})^{c}\tau_{R}^{-}$ is
estimated for T $\gtrsim M_{1}^{'+}$,
\begin{eqnarray}
\Gamma({\phi^{'}}_{1}^{-} \leftrightarrow (\nu^{e}_{L})^{c} \tau_{R}^{-} ) \simeq
 \frac{1}{22\pi}\biggl|\frac{{\gamma}_{2}^{e \tau}}{\sin\beta}\biggr|^{2}\frac{(M_{1}^{'+})^{2}}{T},  \label{phidecay}
\end{eqnarray}
where $M_{1}^{'+}$ is the mass of the field ${\phi^{'}}_{1}^{-}$.
 Similar results are obtained for the processes induced by the coupling constant ${\gamma}_{2}^{e \mu}$.

\subsection{Out-of-equilibrium condition to avoid the baryon number washing out \label{ooec}}

In the present section, we consider which processes discussed in the
previous section were out of equilibrium in the early
Universe \footnote{Strictly speaking, scattering processes like
$\mu^-\nu_{\tau} \leftrightarrow \phi^{0c}_1 \phi^-_2$ may be in
equilibrium for very large $\lambda$.  We do not consider
such a case, however.}.  We will consider what numbers are conserved in the
case where these processes are out of equilibrium in the next section.
The condition that the processes are out of equilibrium is given by
\begin{eqnarray}
\Gamma &<& H,  \label{oe}
\end{eqnarray}
where H is the Hubble parameter, $H=1.66
\sqrt{g_{*}} T^2/M_{Pl} \simeq 1.47 \times 10^{-18}
(T^2/\mbox{GeV})$ with $M_{Pl}$ being the Planck mass.  Here,
$g_{*}$ is the total degrees of freedom of effectively massless
particles.  In the GZM, we adopt the value $g_{*} = g_{*}^{SM} +
g_{*}^{h^{-}} + g_{*}^{\Phi_{2}} \simeq$ 112.75.

\begin{center}1.  Coupling constants $f^{e \mu}, f^{e \tau}$, and $f^{\mu \tau}$ are out of
equilibrium \end{center}
\par
$f^{\mu \tau}$ process. We examine the condition that 
the process $h^{-} \leftrightarrow \mu^{-} \nu_{\tau}$ is out of equilibrium at $T=M_{h}$.
We apply the condition (\ref{oe}) to the rate (\ref{hdecay}),
\begin{eqnarray}
\Gamma(h^{-} \leftrightarrow \mu^{-} \nu_{\tau}) &<& H  \\
\Longleftrightarrow 
 \frac{1}{5.4\pi}|f^{\mu \tau}|^{2}\frac{M_{h}^{2}}{T}  &<& 1.5 \times 10^{-18} 
 \biggl(\frac{T^2}{\mbox{GeV}}\biggr). \label{hoe}
\end{eqnarray}
We obtain the condition for $T=M_{h}$, 
\begin{eqnarray}
2.9 \times 10^{6}\frac{\tan^{4} \beta}{\sin^{4}2 \alpha} \ \mbox{GeV}^{2} 
< \biggr|\log \frac{M_{1}^{2}}{M_{2}^{2}}\biggl|^{4}[M_{1}^{2}+M_{2}^{2}+
(M_{1}^{2}-M_{2}^{2})\cos 2\alpha], \label{hdecay2}
\end{eqnarray}    
where we use the following relations:
\begin{eqnarray}
|f^{\mu \tau}| &=& \biggl|\frac{d}{A m_{\tau}^{2}}\biggr| ,\ 
A=\frac{1}{8\sqrt{2}\pi^{2} v \tan \beta} \sin 2\alpha \biggl|\log \frac{M_{1}^{2}}{M_{2}^{2}}\biggr|,\\ 
M_{h}^{2} &=&\frac{M_{1}^{2}+M_{2}^{2}+(M_{1}^{2}-M_{2}^{2})\cos 2\alpha}{2}.
\end{eqnarray}
At the higher temperatures, $T > M_{h}$, the out-of-equilibrium condition Eq. (\ref{hoe}) is more
 easily satisfied. At lower temperatures, $T < M_{h}$, the number density
  of the Zee singlet $h^{-}$ in the thermal bath is so small
   that the $L_{\mu \tau}$-violating processes are negligible.
   It is therefore sufficient to consider the out-of-equilibrium condition at T $\simeq M_{h}$.
We set $\beta=\pi/4$ for simplicity hereafter.
The inequality (\ref{hdecay2}) is written for this value of $\beta$,
\begin{eqnarray}
\frac{2.9 \times 10^{6}}{\sin^{4}2 \alpha} \ \mbox{GeV}^{2} <  \Biggl(\log \frac{M_{1}^{2}}{M_{2}^{2}}\Biggr)^{4}
\Bigl[M_{1}^{2}+M_{2}^{2}+(M_{1}^{2}-M_{2}^{2})\cos 2\alpha \Bigr]. \label{hdecay3}
\end{eqnarray}
From Eq. (\ref{hdecay3}), we find that this condition is easily satisfied for the angle $\sin 2\alpha \sim 1$.
We then fix the value of $\alpha$ to $\pi/4$.
In this case, the inequality (\ref{hdecay3}) is simplified as 
\begin{eqnarray}
 18 < \Bigl(\log \frac{y}{x} \Bigr)^{4}[x^2+y^2],
\end{eqnarray}
where we parametrize as $x \equiv M_{1}/100$ GeV, $y \equiv M_{2}/100$ GeV ($x,y \geq 1
, x \geq y$). From this, we find that this process is out of equilibrium for $M_{1} \geq 310$ \mbox{GeV},
 when we fix the value of $M_{2}$ to 100 GeV
  \footnote{In Ref. \cite{Haba}, it is difficult for the LMA MSW solution to satisfy the out-of-equilibrium condition
 because the value of the angle $\alpha$ is set to 0.01 rad.}.
\par

$f^{e \tau}$ process. The out-of-equilibrium condition for the $f^{e \tau}$ process is obtained just as
 the condition for the $f^{\mu \tau}$ process. 
We examine the condition that the process $h^{-} \leftrightarrow  e^{-} \nu_{\tau}$ is out of equilibrium at $T=M_{h}$.
We apply the condition (\ref{oe}) to the rate (\ref{hdecay}),
 \begin{eqnarray} 
 \frac{1}{5.4\pi}|f^{e \tau}|^{2}\frac{M_{h}^{2}}{T}  &<& 1.47 \times 10^{-18} \biggl(\frac{T^2}{\mbox{GeV}}\biggr), \\
 \Longleftrightarrow  \frac{2.9 \times 10^{7}}{\sin^{4}2 \alpha} \ \mbox{GeV}^{2} &<&  
 \Biggl(\log \frac{M_{1}^{2}}{M_{2}^{2}}\Biggr)^{4}
\Bigl[M_{1}^{2}+M_{2}^{2}+(M_{1}^{2}-M_{2}^{2})\cos 2\alpha \Bigr]. \label{hdecay4}
\end{eqnarray}
In the case $\alpha=\pi/4$, the inequality (\ref{hdecay4}) is simplified as 
\begin{eqnarray}
 1.8 \times 10^{2} < \Bigl(\log \frac{y}{x}\Bigr)^{4}[x^2+y^2]. \label{etau}
\end{eqnarray}
From this, we find that this process is out of equilibrium for $M_{1} \geq 500$ \mbox{GeV},
 when we fix the value of $M_{2}$ to 100 GeV. 
\par

$f^{e \mu}$ process. The out-of-equilibrium condition for the $f^{e \mu}$ process is obtained in a 
similar manner to that in the $f^{\mu \tau}$ process.
The condition $\Gamma(h^{-} \leftrightarrow e^{-} \nu_{\mu}) < H$ is written as
 \begin{eqnarray}
\frac{2.0 \times 10^{17}}{\sin^{4}2 \alpha} \ \mbox{GeV}^{2} &<&
 \Biggl(\log \frac{M_{1}^{2}}{M_{2}^{2}}\Biggr)^{4}
\Bigl[M_{1}^{2}+M_{2}^{2}+(M_{1}^{2}-M_{2}^{2})\cos 2\alpha \Bigr].
\end{eqnarray}
From this, we find that this process is out of equilibrium for $M_{1}
 \geq 1.2 \times 10^{6}$ \mbox{GeV}, when we fix the value of
 $\alpha$ to $\pi/4$ and the value of $M_{2}$ to 100 GeV.  Since we
 recognize that this mass range for $M_{1}$ is quite large and
 somewhat unnatural, we consider that this process is in equilibrium
 at $T=M_{h}$.

  \begin{center}2. Coupling constants ${\gamma}_{2}^{e \tau}$ and ${\gamma}_{2}^{e \mu}$ are 
 out of equilibrium \end{center}
 \par
 We consider which processes induced from the coupling constants ${\gamma}_{2}^{e \tau}$
 and ${\gamma}_{2}^{e \mu}$ 
 are out of equilibrium for each case considered in Sec. \ref{match}: ${\gamma}_{2}^{e \tau} 
 \not=0$,${\gamma}_{2}^{e \mu}=0$ and 
 ${\gamma}_{2}^{e \tau}=0$,${\gamma}_{2}^{e \mu} \not=0$.
\par

${\gamma}_{2}^{e \tau} \not=0$ and ${\gamma}_{2}^{e \mu}=0$. We examine the condition that 
the process ${\phi^{'}}_{1}^{-} \leftrightarrow (\nu^{e}_{L})^{c} \tau_{R}^{-} $ is out of equilibrium at $T=M_{h}$,
 \begin{eqnarray}
&&\Gamma({\phi^{'}}_{1}^{-} \leftrightarrow (\nu^{e}_{L})^{c} \tau_{R}^{-} )  < H,  \\
&&\Rightarrow \ 8.4 \times 10^{10} \ \mbox{GeV}  < \sqrt{\frac{M_{1}^2+M_{2}^2}{2}},  
 \end{eqnarray}
 where we fix the value of $\alpha$ to $\pi/4$. 
 From this, we find that this process is out of equilibrium for $M_{1} \geq 1.2 \times 10^{11} \mbox{GeV}$,
 when we fix the value of $M_{2}$ to 100 GeV.
  We consider that this process is in equilibrium at $T=M_{h}$ for the same reason mentioned above.
  \par
${\gamma}_{2}^{e \tau}=0$ and ${\gamma}_{2}^{e \mu}\not=0$. We examine the condition that 
the process ${\phi^{'}}_{1}^{-} \leftrightarrow (\nu^{e}_{L})^{c} \mu_{R}^{-} $ is out of equilibrium as follows:
 \begin{eqnarray}
&&\Gamma({\phi^{'}}_{1}^{-} \leftrightarrow (\nu^{e}_{L})^{c} \mu_{R}^{-} )  < H,  \\
&&\Rightarrow  2.8 \times 10^{8} \ \mbox{GeV}  < \sqrt{\frac{M_{1}^2+M_{2}^2}{2}},  
 \end{eqnarray}
  where we fix the value of $\alpha$ to $\pi/4$. 
 From this, we find that this process is out of equilibrium for $M_{1} \geq 4.0 \times 10^{8}$ \mbox{GeV},
 when we fix the value of $M_{2}$ to 100 GeV.
   We consider that this process is also in equilibrium at $T=M_{h}$.

\subsection{Final lepton number and baryon number}
In the present section, we consider the final lepton and  baryon number for the two cases
 ${\gamma}^{e \tau}_{2} \not=0 ,{\gamma}^{e \mu}_{2}=0 $ and 
${\gamma}^{e \tau}_{2} =0 ,{\gamma}^{e \mu}_{2}\not=0$. 

 \begin{center}1. ${\gamma}^{e \tau}_{2} \not=0 ,{\gamma}^{e \mu}_{2}=0$   \end{center}
 \par
\begin{figure}[ht]
\begin{center}
\unitlength=1mm
\begin{picture}(80,80)
\epsfile{file=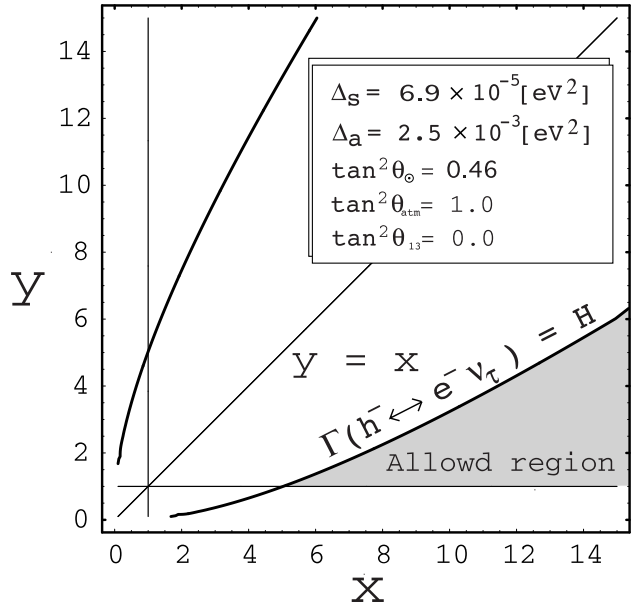,height=8cm}
\end{picture}
\caption{The primordial baryon number remains finite in the allowed
 region\label{phase1}.  The out-of-equilibrium condition $1.8 \times
 10^{2} < [\log (y/x)]^{4}[x^2+y^2]$ [Eq. (\ref{etau})] is used in this figure with the variables 
 $x \equiv M_{1}/100 \ \mbox{GeV} ,\ y \equiv
 M_{2}/100$ GeV.  Only the region $\ x \geq 1,\ y
 \geq 1,\ x \geq y$ is considered.}
\end{center}
\end{figure}

Since the processes induced by ${\gamma}_{2}^{e \tau}$ and $f^{e \mu}$ are in equilibrium from the previous section,
 the only possible conserved quantity is 
 \begin{eqnarray}
 P \equiv L_{e \tau}.
\end{eqnarray} 
This quantity is conserved if the process induced by $f^{e\tau}$ is
out of equilibrium [Eq. (\ref{hdecay4})]. The region in which this
quantity is conserved is shown in Fig. \ref{phase1}.  This conserved
number is converted to the baryon number through the sphaleron process
and the final baryon ($B^{eq}_{f}$) and lepton number ($ L^{eq}_{f}$)
are calculated as 
\begin{eqnarray} 
 B^{eq}_{f}&=&\frac{60}{563}\Bigl(\frac{B}{3}+L_{e \tau}\Bigr)_{i}, \nonumber\\
 L^{eq}_{f}&=&-\frac{39}{20}B^{eq}_{f},
\end{eqnarray} 
where $(B/3+L_{e \tau})_{i}$ is the primordial value generated
at the temperature much higher than $T=M_{h}$.  The region shown in
Fig. \ref{phase1} is therefore the allowed region for avoiding the
washing out of the primordial baryon number in this case.

 \begin{center} 2. ${\gamma}^{e \tau}_{2}=0, {\gamma}^{e \mu}_{2}\not=0$  \end{center}  
 \par
Since the processes induced by ${\gamma}_{2}^{e \mu}$ and $f^{e \mu}$ are in equilibrium,
 the only possible conserved quantity in this case is 
 \begin{eqnarray}
 P \equiv L_{e \tau}+L_{\mu \tau} (=-2L_{\tau}).
\end{eqnarray} 
This quantity is conserved if the process induced from $f^{e\tau}$ and
 $f^{\mu\tau}$ is out of equilibrium [Eq. (\ref{hdecay4}) and
 Eq. (\ref{hdecay3})]. The region in which this quantity is conserved
 is shown in Fig. \ref{phase1}.  This conserved number is converted to
 the baryon number through the sphaleron process and the final baryon
 ($B^{eq}_{f}$) and lepton number ($ L^{eq}_{f}$) are calculated as
 \begin{eqnarray}
 B^{eq}_{f}&=&\frac{60}{185}\Bigl(\frac{B}{3}-L_{\tau}\Bigr)_{i}, \nonumber\\
 L^{eq}_{f}&=&-\frac{39}{20}B^{eq}_{f},
\end{eqnarray} 
where $(B/3-L_{\tau})_{i}$ is a primordial value generated
 at the temperature much higher than $T=M_{h}$.   The region shown in
FIG. \ref{phase1} is therefore the allowed region for avoiding the
washing out of the primordial baryon number in this case, also.

\subsection{WMAP and Z burst}
The investigation into the cosmic microwave background  by the Wilkinson
 Microwave Anisotropy Probe (WMAP) recently gave the
following upper bound to the sum of the neutrino masses
\cite{Wmap1,Wmap2,Wmap3,Wmap4}:
\begin{eqnarray} \sum_{i} |m_{i}| <
0.71\ \mbox{eV} \ (95\% \ \mbox{C.L.}).
\end{eqnarray} 
This value is consistent with that obtained in Sec. \ref{A},
\begin{eqnarray}
  \sum_{i} |m_{i}|=\sqrt{(2a-\epsilon)^2+8b^2}+d \simeq 0.13 \ \mbox{eV}.
\end{eqnarray}
\par

If the ultrahigh energy cosmic rays with energy beyond the
Greisen-Zatsepin-Kuzmin limit are explained in the Z-burst scenario, the
heaviest neutrino mass should lie in the range \cite{Wei,Far,Fod}
\begin{eqnarray} 0.08 \ \mbox{eV} < |m_{\nu}(\mbox{heaviest})| < 1.3
\ \mbox{eV} \ (68 \% \ \mbox{C.L.}).
\end{eqnarray} 
The value obtained in Sec. \ref{A},
\begin{eqnarray}
 |m_{1}|=\frac{\epsilon+\sqrt{(2a-\epsilon)^2+8b^2}}{2} 
\simeq 5.3 \times 10^{-2}  \ \mbox{eV},
\end{eqnarray}
is near this range, and the ultrahigh energy cosmic rays might be
explained in the Z-burst scenario in our model.

\section{Summary}
We propose a simple GZM which is compatible with all the neutrino
oscillation data for the LMA MSW solution of the solar neutrino
problem, and is free from the baryon number washing out.  Our
extension from the RZM is minimal, i.e., we add only one coupling
constant $\gamma_{2}^{e\tau}$ or $\gamma_{2}^{e\mu}$ in addition to
those in the RZM. To avoid the baryon number washing out, the masses
$M_{1}$ and $M_{2}$ must be in the region shown in Fig. \ref{phase1}.
The final baryon and lepton numbers are calculated for the two cases
(1)${\gamma}_{2}^{e \tau} \not=0 ,{\gamma}_{2}^{e \mu}=0$ and (2)
${\gamma}_{2}^{e \tau} =0 ,{\gamma}_{2}^{e \mu}\not=0$:
\begin{eqnarray} &(1)& {\gamma}_{2}^{e \tau} \not=0 ,{\gamma}_{2}^{e
\mu}=0:B^{eq}_{f} =\frac{60}{563}\biggl(\frac{B}{3}+L_{e \tau}\biggr)_{i} ,\
L^{eq}_{f} = -\frac{39}{20}B^{eq}_{f}, \\ &(2)& {\gamma}_{2}^{e \tau}
=0 ,{\gamma}_{2}^{e \mu} \not=0:B^{eq}_{f}
=\frac{60}{185}\biggl(\frac{B}{3}-L_{\tau}\biggr)_{i} ,\ L^{eq}_{f} =
-\frac{39}{20}B^{eq}_{f}.
\end{eqnarray} 
In order to explain today's baryon number, it is therefore necessary
that the quantity $(B/3+L_{e \tau})$ or
$(B/3-L_{\tau})$ is generated in the early Universe or the
baryon number should be produced at lower temperatures  $T \ll
M_{h}$. We check that the predicted lepton flavor violating processes
are not in conflict with the phenomenological constraints for
$\tau(\mu)$ decay, $\tau(\mu) \rightarrow e \gamma$, and 0$\nu\beta
\beta$ decay.

\begin{acknowledgments}
We thank N. Haba for fruitful discussions and for providing us
with the detailed calculation of Ref. \cite{Haba}.  One of the author
(K.O.) thanks the Japan society for the promotion of
science for financial support (No.4834). The work of C.S.L. was supported in part by a
Grant-in-Aid for Scientific Research of the Ministry of Education,
Science, and Culture, Grant No.80201870.
\end{acknowledgments}

\end{document}